\documentclass[conference,letterpaper]{IEEEtran}
\IEEEoverridecommandlockouts
\usepackage{cite}
\usepackage{amsmath,amssymb,amsfonts,bm}
\usepackage{algpseudocode}
\usepackage{algorithm,float}
\usepackage{graphicx,subfigure}
\usepackage{textcomp}
\usepackage{xcolor}
\usepackage{multirow}
\usepackage{balance} 
\usepackage{colortbl}
\usepackage{url}
\usepackage{caption}
\usepackage{graphicx,epstopdf}
\usepackage{grffile}
\def\BibTeX{{\rm B\kern-.05em{\sc i\kern-.025em b}\kern-.08em
T\kern-.1667em\lower.7ex\hbox{E}\kern-.125emX}}

\definecolor{LightCyan}{rgb}{0.88,1,1}
\definecolor{LightGray}{rgb}{0.88,0.88,0.88}

\definecolor{org}{rgb}{0.9,0.4,0.1}

\newcommand\Tstrut{\rule{0pt}{2.6ex}}         

\begin{document}
\title{
	Interference-Aware Super-Constellation \\  Design for  NOMA
}

\author{%
	\IEEEauthorblockN{Mojtaba Vaezi$^{\dagger}$ and Xinliang Zhang$^{\ddagger}$  }
	\IEEEauthorblockA{
		$^{\dagger}$Department of ECE,
		Villanova University, Villanova, PA 19085, USA\\
				$^{\ddagger}$Samsung Research America, 6625 Excellence Way, Plano, $\rm Tx$ 75023, USA  \\
		Emails:   mvaezi@villanova.edu$^\ddagger$, leon.zhang@ieee.org$^\dagger$
	}
	\thanks{This work was supported by the by the U.S. National Science Foundation (NSF) under Grant ECCS-2301778.
}
}

\maketitle

\begin{abstract}
	Non-orthogonal multiple access (NOMA) has gained significant attention as a potential next-generation multiple access technique. However, its implementation with finite-alphabet inputs faces challenges. Particularly, due to inter-user interference,  superimposed constellations may have overlapping symbols leading to high bit error rates when successive interference cancellation (SIC) is applied. To tackle the issue, this paper employs  autoencoders to design interference-aware super-constellations.  Unlike conventional methods where superimposed constellation may have overlapping symbols, the proposed autoencoder-based NOMA (AE-NOMA) is trained to design super-constellations with distinguishable symbols at receivers, regardless of channel gains.  
The proposed architecture removes the need for SIC, allowing maximum likelihood-based approaches to be used instead. The paper presents the conceptual architecture, loss functions, and training strategies for AE-NOMA. Various test results are provided to demonstrate the effectiveness of interference-aware constellations in improving the bit error rate, indicating the adaptability of AE-NOMA to different channel scenarios and its promising potential for implementing NOMA systems.
\end{abstract}


\section{Introduction}


Non-orthogonal multiple access (NOMA) has attracted significant academic interest as a next-generation multiple access candidate \cite{Jafarkhani2024noma}. While NOMA has also been explored by 3GPP \cite{NOMA3GPPNR}, it has not been adopted into standards, primarily due to the gap between theoretical and experimental performance \cite{vanka2012superposition,qi2021over,vaezi2019multiple}. A main challenge is implementing successive interference cancellation (SIC) to address inter-user interference. Specifically, with finite-alphabet inputs, like quadrature phase-shift keying (QPSK) constellations, superimposing two constellations may result in a \textit{super-constellation} with overlapping or closely spaced symbols \cite{cejudo2017power,kara2018ber,han2021study,Jafarkhani2024noma,assaf2020exact}, leading to high decoding error rates in the symbol and bit levels.  The issue stems from the fact that traditional constellations are designed for point-to-point channels and  their constellation shaping is oblivious to interference. In contrast, through superposition, NOMA inherently incorporates inter-user interference, indicating that using traditional constellations in NOMA channels can introduce limitations and mismatches depending on users' channel gains.

Alternatively, NOMA constellations can be designed using an autoencoder (AE) \cite{Jafarkhani2024noma}. The idea of employing AEs for communication was first introduced in \cite{o2017introduction} and later refined and extended to various domains including interference management \cite{nartasilpa2018communications,zhang2021svd,zhang2023deep,chahine2023deepic+}, uplink NOMA  \cite{ye2020deepnoma,ye2021deep}, and downlink NOMA  \cite{alberge2018constellation, van2022deep, ninkovic2023weighted, van2020deep}.  
The above AE-NOMA studies collectively aim to implement SIC, either structurally or through their loss functions, following the principles of infinite-alphabet NOMA.

Unlike the aforementioned works, in this paper we introduce SIC-free AE-NOMA to generate super-constellations that are inherently resilient to inter-user interference rather than trying to remove interference using SIC. The proposed AE-NOMA designs super-constellations with distinguishable symbols, independent of NOMA user channel gains. This contrasts with the conventional approach, where each user is assigned a predefined constellation (e.g., QPSK), and their superposition forms the super-constellation. With this conventional approach, depending on the percentage  of power allocated to each NOMA user, some symbols of super-constellation may overlap or become very close to others, which  limits the use of NOMA. Unlike that, our approach is aware of the interference in the design phase and will result in super-constellations with a good minimum distance. More importantly, our design realizes SIC-free decoding which is crucial for resource-limited devices. 

The contributions of the paper can be summarized as 
\begin{itemize}
	\item We introduce  autoencoder-based interference-aware constellation design for NOMA by designing and optimizing the end-to-end system to craft well-distinguishable super-constellations. 
 \item We design an adaptive weighted loss function to optimize network training to enable the adjustment of weights for improved user fairness based on bit error rates (BERs).

\item We conduct  extensive testing to gain insights into how interference-aware constellations could enhance the BER through the implementation of SIC-free NOMA.
		
\end{itemize}

\begin{figure*}[htbp] 
	\centering
	\includegraphics[scale=.61]{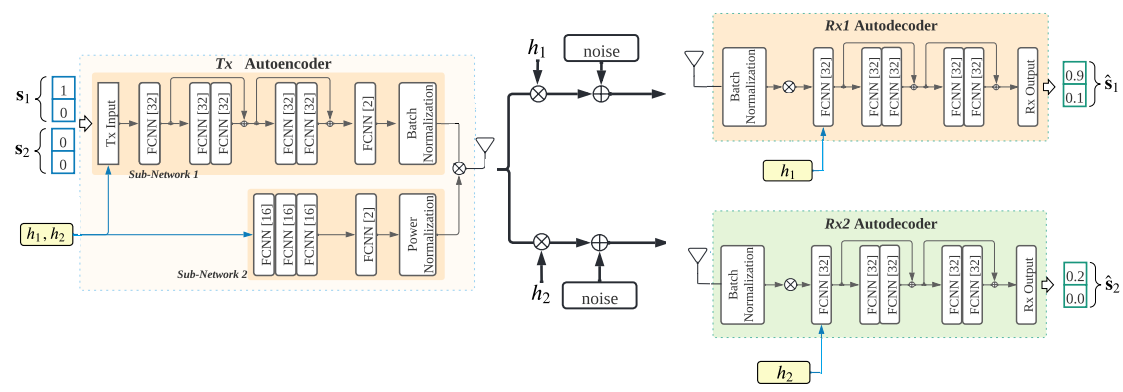}
	\caption{The architecture of the implemented AE-NOMA for two-users. Each FCNN[$\ell$] represents a fully connected layer with $\ell$ nodes. The main structure which consists five hidden layers each with 32 neurons is replicated at $\rm Tx$, $\rm Rx1$ and $\rm Rx2$. The $\rm Tx$, however, has a side block (Sub-Network 2) which is used to adjust power allocation for each symbol. }
	\label{fig_DAE_net} 
\end{figure*}

 The resulting BERs in this work are significantly better than those in \cite{alberge2018constellation, van2022deep, ninkovic2023weighted, van2020deep} across all ranges of signal-to-noise ratios (SNRs).
 For instance, the work in \cite{ninkovic2023weighted}, which enhances the results of \cite{alberge2018constellation}, still yields BERs that are about an order of magnitude worse than our results. A detailed discussion and comparison of these results along with the assumptions  are provided in our numerical findings.

\section{System Model}
Consider a downlink NOMA system involving a base station (BS) communicating with two users. All nodes are equipped with single antennas.  The user with weaker (stronger) channel is referred to as the weak (strong) user. The channels between the BS and the weak and strong users are respectively denoted by $h_1$   and $h_2$ satisfying $|h_1|^2 \le |h_2|^2$. Both channel gains are known at the BS, but each user only knows its own channel gain. All channel gains are complex-valued and independently drawn from a continuous distribution.

 In a NOMA system, the BS employs superposition coding to concurrently transmit messages to both users. The transmitted signal $x$ is defined by $x = \sqrt{\alpha P}s_1 +\sqrt{\bar \alpha P}s_2$, where $s_1$ and $s_2$ are independent and identically distributed complex Gaussian signals with zero mean and unit variance,  $\mathcal{CN}(0, 1)$. Here, $P$ represents the BS's transmit power budget, $\alpha \in [0, 1] $ is the fraction of power allocated to the weak user, and $\bar \alpha \triangleq 1 - \alpha$. The received signals at the users are expressed as
\begin{align}
y_k &= h_k x + n_k, \qquad k\in\{1,2\}
\end{align} 
where the noises $n_1$ and $n_2$ are independent and identically distributed circularly-symmetric complex Gaussian random variables with zero mean and unit variance, $\mathcal{CN}(0, 1)$.

The weak user decodes its message by treating the interfering signal from the other user as noise. In contrast, the strong user first decodes the weak user’s message, treating its own interfering signal as noise, and then applies SIC to decode its own message. The combined use of superposition coding and SIC decoding with Gaussian codebooks results in achieving the capacity region of this channel.

 \section{Network Structure}

\subsection{Autoencoder's Structure}

   An autoencoder  consists of two parts, an encoder and a decoder.
   The AE processes a binary input  vector $\mathbf{s} \in [0, 1]^d$ by first mapping it to a hidden representation $\mathbf{u} \in [0, 1]^{d'}$ through a deterministic mapping $\mathbf{u} = f_\theta(\mathbf{s}) = \sigma(W\mathbf{s} + \mathbf{b})$, where $\theta = \{W, \mathbf{b}\}$, $W$ is a $d' \times d$ weight matrix, and $\mathbf{b}$ is a bias vector. Here, $\sigma(x) = \frac{1}{1 + e^x}$ is the sigmoid  function and $\sigma(\mathbf{x}) = [\sigma(\mathbf{x}_1), \dots, \sigma(\mathbf{x}_d)]^T$. The resulting latent representation $\mathbf{u}$ is then transformed back to a reconstructed vector $\hat{\mathbf{s}} \in [0, 1]^d$ in the input space, given by $\hat{\mathbf{s}} = g_{\theta'}(\mathbf{u}) = \sigma(W'\mathbf{u} + \mathbf{b}')$ with $\theta' = \{W', \mathbf{b}'\}$. 
   For each training sample $\mathbf{s}^{(i)}$, the AE maps it to a corresponding $\mathbf{u}^{(i)}$ and a reconstruction $\hat {\mathbf{s}}^{(i)}$.

   
 The proposed AE-NOMA network, illustrated in Fig.~\ref{fig_DAE_net}, consists of one transmitter  ($\rm Tx$) encoder and two decoders each at one of the receivers, i.e., $\rm Rx1$ and $\rm Rx2$. Each encoder and decoder, include a fully connected neural network (FCNN) composed of a sequence of fully connected layers and some residual  connections.  
 As  seen in Fig.~\ref{fig_DAE_net}, the encoder ($\rm Tx$) comprises a main block and a secondary network, referred to as Sub-Network 1 and Sub-Network 2, respectively. 
 
 This main block, replicated at $\rm Tx$, $\rm Rx1$ and $\rm Rx2$, consists of an input layer, five hidden layers, and an output layer.   The hidden layers contain 32 neurons each, while the final layer (output layer) has two neurons representing the in-phase (I) and quadrature-phase (Q) components of the symbol designed by the autoencoder for transmission. In addition to the main block, the $\rm Tx$ has a side block (Sub-network 2) which helps to better adjust the I and Q components under a given average power constraint, thereby optimizing the use of the I/Q plane in a way similar to a quadrature amplitude modulation (QAM).
 
  In the main block, we have also implemented residual connections to enhance learning capacity and improves performance without the need for additional parameters or a wider network. These shortcuts also play a crucial role in preserving gradients within the network. We have used the activation function \textit{exponential linear unit} which provides a smooth, non-zero output for negative inputs, and can help improve learning stability and performance.

\subsection{Loss Function}

Each output vector in the two receivers represents binary messages, then the network can be trained using binary cross-entropy loss $	\mathcal{L}_k$  for user $k\in\{1,2\}$ which is given by
\begin{align}
	\mathcal{L}_k=\frac{1}{N_B}\sum_{n=1}^{N_B}    (\mathbf{s}_{k,n})^T\log\hat{\mathbf{s}}_{k,n} +
	(1-\mathbf{s}_{k,n})^T\log(1-\hat{\mathbf{s}}_{k,n}),
\end{align}
where  $N_B$ is the batch size, $\mathbf{s}_{k,n}$ is the 
$n$th  
input bit-vector in the batch, and $\hat{\mathbf{s}}_{k,n}$ is the corresponding output (the predicted probability for each element of the input vector obtained from a softmax output layer).  Note that $\mathbf{s}_{k,n}$ is a column vector and $\log\hat{\mathbf{s}}_{k,n}$ involves applying the logarithm element-wise to each element  $\hat{\mathbf{s}}_{k,n}$.  The 
loss function treats each element of the AE output as a 0/1 classification task. The binary cross-entropy loss function is commonly used for multi-label classification problems, where each example can have multiple binary labels. The loss function measures the difference between the predicted probability of each label being present in the example and the true probability of the label being present.  
In the training process, the backpropagation algorithm passes $\mathcal{L}_1$ to \textit{$\rm Rx1$} 
and it will further go to \textit{$\rm Tx$}  whereas   $\mathcal{L}_2$  affects
\textit{$\rm Rx2$} and \textit{$\rm Tx$}.
In our AE-NOMA, each receiver has its own estimation of the transmitted bits. Then, we define the  overall 
loss function is defined as 
\begin{align} \label{L}
	\mathcal{L}=w_1\mathcal{L}_1+w_2\mathcal{L}_2,
\end{align}
\noindent where $w_1$ and $w_2$ are \textit{loss weights} to adjust the loss; $\mathcal{L}_1$ and $\mathcal{L}_2$ are the loss  at \textit{$\rm Rx1$} and \textit{$\rm Rx2$}. We propose the following symmetric weights  
\[
\begin{cases}
	w_1=w,\; w_2 = 1, & \text{if } \mathcal{L}_1 \ge \mathcal{L}_2 \\
	w_1=1,\; w_2 = w, & \text{if } \mathcal{L}_1 < \mathcal{L}_2 \\
\end{cases}
\] which are \textit{adaptive} to $\mathcal{L}_1$ and $\mathcal{L}_2$.
The \textit{loss weight} $w$ is a coefficient set in the training process. 
\begin{align}
&\theta^\star, \theta_1'^\star, \theta_2'^\star = \operatorname*{arg\,min}_{\theta,\theta_1', \theta_2'} \; \mathcal{L}   \\  &=  \operatorname*{arg\,min}_{\theta,\theta_1', \theta_2'} \; w_1\mathcal{L}_1 (\mathbf{s}_1, g_{\theta_1'}( f_{\theta}(\mathbf{s}_1))) + w_2   \mathcal{L}_2 (\mathbf{s}_2, g_{\theta_2'}( f_{\theta}(\mathbf{s}_2))) \notag
\end{align}
where $\theta,\theta_1'$, and $\theta_2'$ are the weights of the $\rm Tx$, $\rm Rx1$, and $\rm Rx2$, respectively, and the optimization is carried out by stochastic gradient descent with Adam optimizer. 

\section{Numerical Results}
The performance of the AE-NOMA is evaluated in this section. We consider two different training scenarios. 
\subsection{Trained with a Fixed Set of Channels  (Test Case 1)}

\begin{figure}[t] 
	\centering
	\includegraphics[scale=0.49, trim=0 0 0 0, clip]{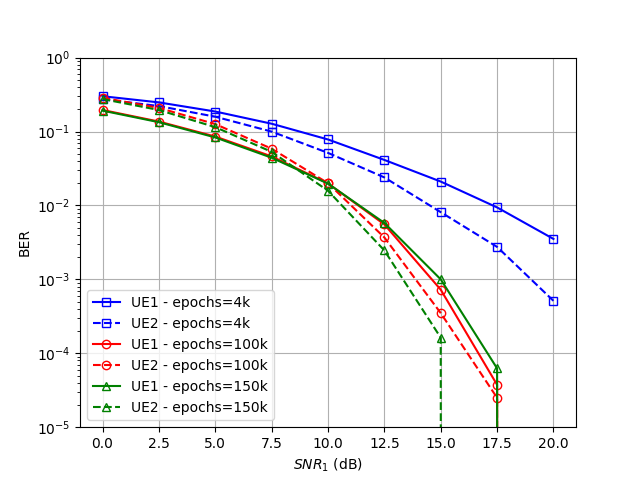}
	\caption{BERs versus  SNR$_1$ for $h_1=1$, $h_2=2$. and $w =10$ for different number of epochs. }
	\label{fig_BER_SNR} 
\end{figure}

\begin{figure}[!h] 
	\centering
%
%
	\subfigure[Constellation at UE1 ($h_1=1$ and   SNR$_1=10\, \rm dB$)]{\centering
		\includegraphics[scale=0.45, trim=0 0 0 0, clip]{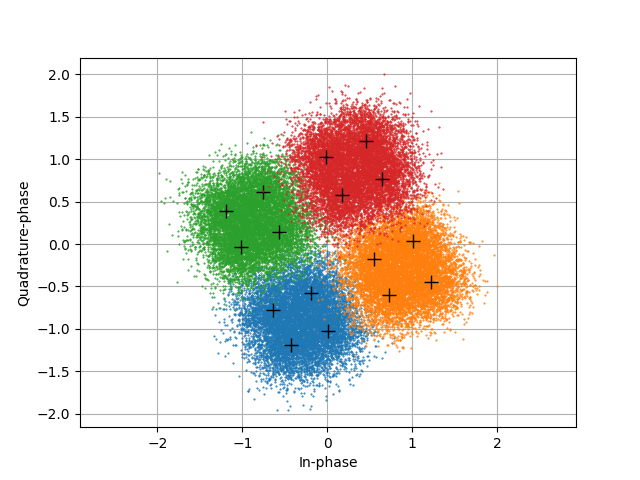}
		\label{fig_BER_constUE1} 	}\\
	
	\subfigure[Constellation at UE2 ($h_2=2$ and   SNR$_2=16\, \rm dB$)]{\centering
			\includegraphics[scale=0.45, trim=0 0 0 0, clip]{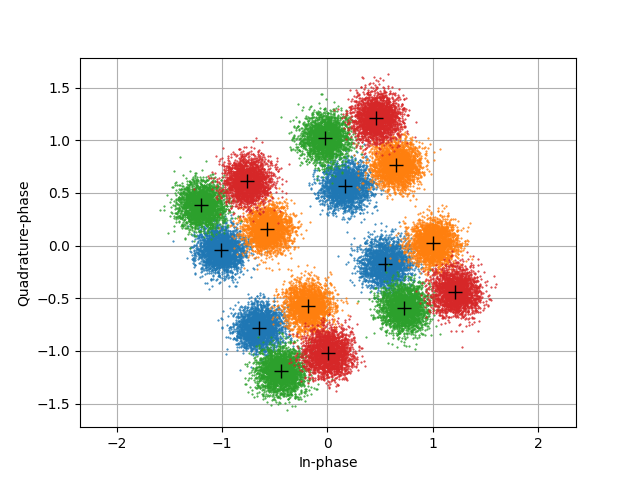}		
		\label{fig_BER_constUE2}}
	
	\caption{AE-NOMA constellations at UE~1 and UE~2  when  $h_2=2h_1 = 2$ were fixed during training and test (Test Case~1), and the number of epochs is 150K. The `+' signs represent noiseless super-constellation symbols ($\rm Tx$ output), whereas the dots represent noisy versions. Each color represents two bits.}
	\label{fig_BER_const}
\end{figure}

The AE-NOMA network is trained with parameters $h_1=1$, $h_2=2$, SNR$_1=10\, \rm dB$ (SNR for $\rm Rx1$), and $w=10$. The noise power at both receivers is identical. The network is tested with the same channel gains ($h_1=1$, $h_2=2$), for which BERs are displayed in Fig.~\ref{fig_BER_SNR} for different number of epochs. The constellations for the two users are shown in Fig.~\ref{fig_BER_const}, with SNR$_1=10\,\rm dB$ and SNR$_2=16\,\rm dB$ (given $h_1=1, h_2=2$).  The black `+' signs indicate the designed (noiseless) super-constellation symbols at the $\rm Tx$. Each dot represents a received symbol, and symbols with the same message to the user share the same color.   As seen in Fig.~\ref{fig_BER_constUE1}, the weaker user (UE1) can distinguish only four clusters, whereas the stronger user (UE2) can differentiate 16 clusters. Consequently, UE1 decodes 2 bits of the message, while UE2 decodes 4 bits—two for UE1 and two for itself.

%

\subsection{Trained with Randomly-Generated Channels}
%

In this subsection, $P=1$, SNR$_1=10  \, \rm dB$, $h_1=1$, and $w=10$ are fixed, similar to the previous subsection. However, in contrast,  the network is trained with randomly 
generated $h_2$ which is uniformly distributed in $[h_{\min}, h_{\max}]$ and tested for one instance of  $h_2$ in the same range. The performance of the AE-NOMA is evaluated in this section.

\subsubsection{Test Case 2}\label{sec_h2_2}
\begin{figure}[t] 
	\centering
	\vspace{.55cm}
	\includegraphics[scale=0.5, trim=0 0 0 0, clip]{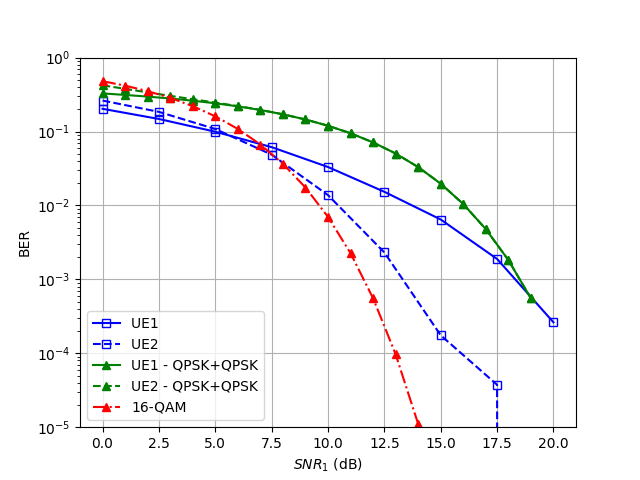}
	\caption{BERs versus  SNR$_1$ for $h_2=2h_1=2$ for AE-NOMA  compared to the theoretical BER for when both users use a QPSK constellation, and BER of a 16-QAM constellation.   }
	\label{fig_BER_SNRV} 
\end{figure}

 The AE-NOMA  is trained with  $h_{\min}=1$ and $h_{\max}=3$, i.e.,  $h_2\sim {\rm Unif}[1,3]$, and an identical noise power at both receivers. Testing results with $h_1=1$ and $h_2=2$  is depicted in Fig.~\ref{fig_BER_SNRV}. 
 In this graph, we present the BER for both users as achieved by AE-NOMA, alongside the optimal theoretical BER when both users utilize a QPSK constellation, with power allocation optimized to satisfy \(\text{BER}_1 = \text{BER}_2\). The BER formulas are detailed in \cite{cejudo2017power}. Through a straightforward optimization, it is shown that setting \(\alpha = 0.7\) ensures \(\text{BER}_1 = \text{BER}_2\); therefore, this value has been used in the graph above.
 
Additionally, we have plotted  the probability of error for a 16-QAM constellation \cite{proakis2008digital}, for   $h_2=2h_1=2$. This choice is motivated by the observation that, from UE2's perspective, an ideal super-constellation should resemble a 16-QAM structure (or its rotations), as this configuration offers the best minimum distance,\footnote{However, this does not necessarily hold from UE1's perspective, except when \( h_1 = h_2 \), as can be inferred from the constellations in Fig.~\ref{fig_BER_const}. When \( h_1 < h_2 \), there should be spacing between every four symbols of the 16-QAM constellation, like in Fig.~\ref{fig_BER_constUE2}. This ensures that \( \text{BER}_1 \) remains low.
} and may be seen as the theoretical lower bound.

\begin{figure}[!htbp] 
	\centering
	\subfigure[Constellation at UE1 ($h_1=1$ and   SNR$_1=10\, \rm dB$)]{\centering
		\includegraphics[scale=0.45, trim=0 0 0 0, clip]{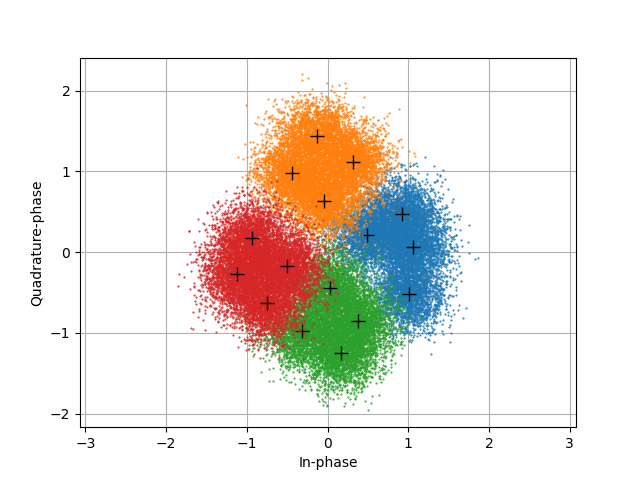}}\\
	
	\label{fig_BER_constUE1V} 
	\subfigure[Constellation at UE2 ($h_2=2$ and   SNR$_2=16\, \rm dB$)]{\centering
		\includegraphics[scale=0.45, trim=0 0 0 0, clip]{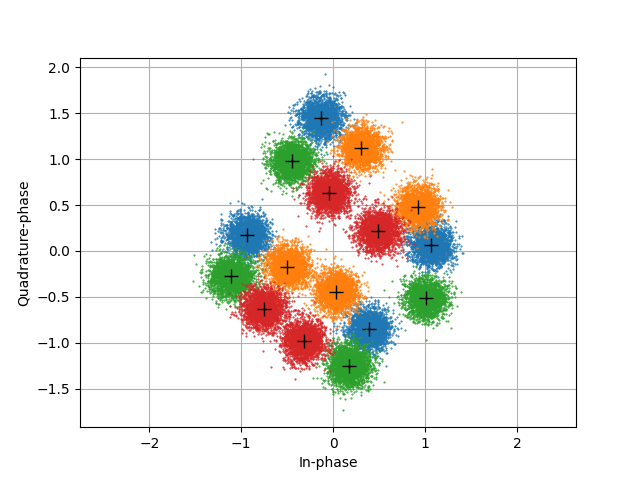}}
	\label{fig_BER_constUE2V} 
	\caption{Constellations at both users for  $h_2=2h_1=2$. Here,  during training  $h_2\sim {\rm Unif}[1,3]$ (Test Case~2). }
	\label{fig_BER_constV}
\end{figure}

We observe that the BER of the constellation designed by AE-NOMA for UE2 nearly follows that of 16-QAM. Interestingly, for \({\text{SNR}}_1 \in [0, 7] \,  \rm dB\), AE-NOMA-designed constellations outperform 16-QAM in terms of BER.  
The AE-NOMA constellations for both users are shown in Fig.~\ref{fig_BER_constV}, with UE1 having \(\text{SNR}_1 = 10\, \rm dB\) and UE2 having \(\text{SNR}_2 = 16\, \rm dB\) (\( h_1 = 1 \) and \( h_2 = 2 \)).  
In Test Case~2, the number of training epochs was 150K, and \( w = 20 \).

%
%
%
%
%
%



\subsubsection{Test Case 3}\label{sec_h2_10}
\begin{figure}[t] 
	\centering
	\includegraphics[scale=0.5, trim=0 0 0 0, clip]{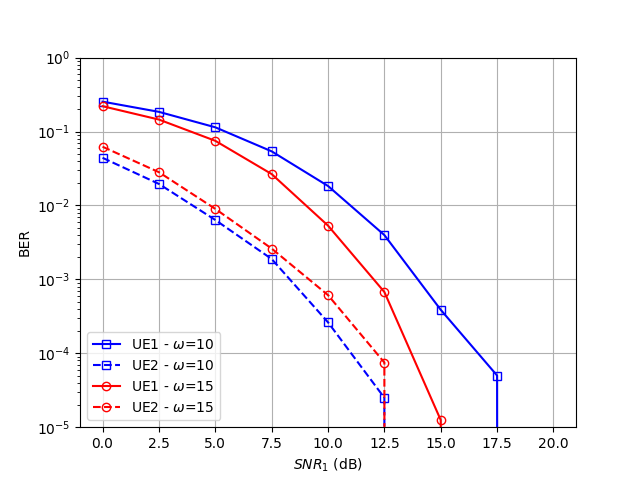}
	\caption{BERs versus  SNR$_1$ for $h_1=1$ and $h_2=10$.}
	\label{fig_BER_SNRV_h2_10} 
\end{figure}

\begin{figure}[h] 
	\centering
	\subfigure[Constellation at UE1 ($h_1=1$ and   SNR$_1=10\, \rm dB$)]{\centering
		\includegraphics[scale=0.45, trim=0 0 0 0, clip]{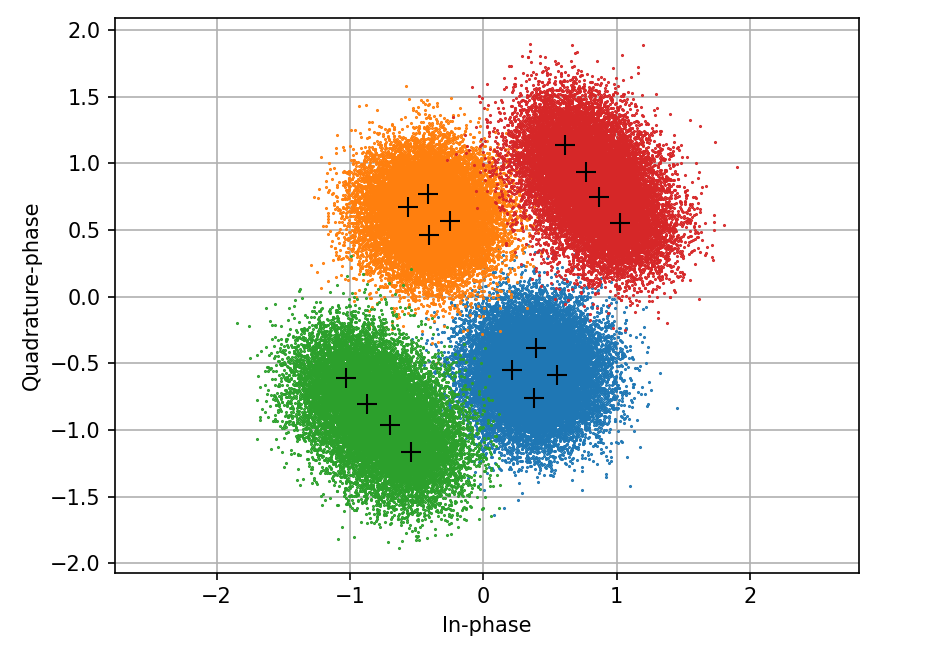}}\\
	
	\label{fig_BER_constUE1V_h2_10} 
	\subfigure[Constellation at UE2 ($h_2=10$ and   SNR$_2=30\, \rm dB$)]{\centering
		\includegraphics[scale=0.45, trim=0 0 0 0, clip]{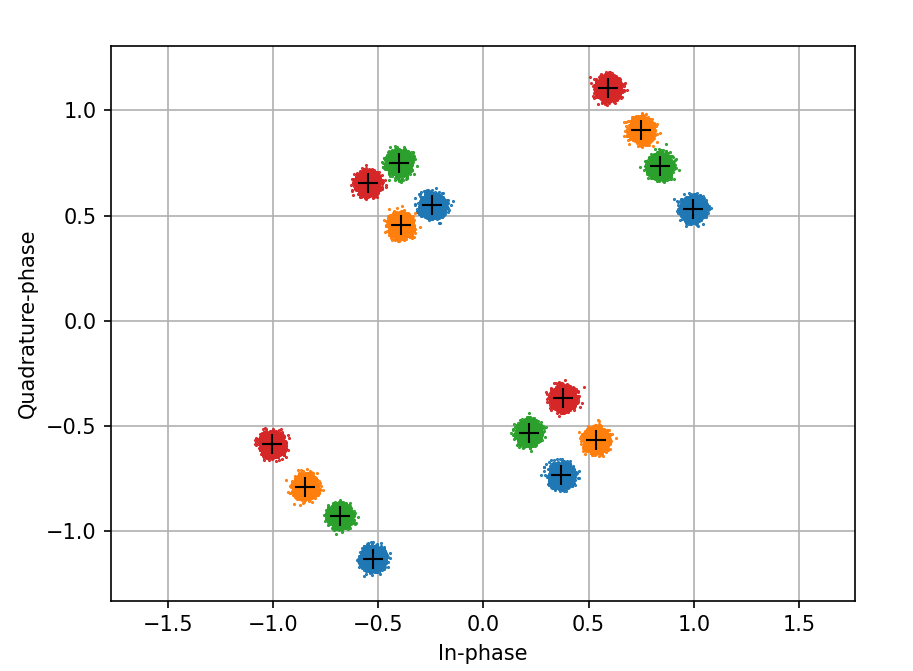}}
	\label{fig_BER_constUE2V_h2_10} 
	\caption{Constellations at both users for $h_1=1$, $h_2=10$. Here,  during training  $h_2\sim {\rm Unif}[8,12]$ (Test Case~3).}
	\label{fig_BER_constV_h2_10}
\end{figure}

In this scenario, the AE-NOMA network is trained with  $h_2\sim {\rm Unif}[8,12]$ during the training phase, while $h_2=10$ is used for the testing phase. This  presents a more challenging test, given the order of magnitude difference between the channel gains ($h_1=1$ and $h_2=10$). The BER results are plotted in Fig.~\ref{fig_BER_SNRV_h2_10} for $w=10$ and $w=15$, indicating that the loss weight $w$ is another parameter to be optimized. The corresponding constellations of the two users are shown in Fig.~\ref{fig_BER_constV_h2_10}, where \( w = 15 \) and the number of epochs is 150K.


\subsection{Analyzing the  Results}\label{sec_h2_10_com}
First, we compare  Test Case~1 and  Test Case~2. Although the plots are for the same set of parameters, namely $h_1=1$ and $h_2=2$, the results, including BER curves and constellations, exhibit noticeable differences. This is because the two cases are trained for different scenarios. Test Case~2, trained with varying values of $h_2$,  yielding to slightly worse BER results compared to Test Case~1. 
%
%
Expanding the range of $h_2$ to introduce higher uncertainty further worsens the results. Another notable observation is the high adaptability of AE-NOMA in constellation design, where even slight changes in the scenario can lead to significantly different constellation shapes.

Next, let us compare Test Case 2 and Test Case 3. We note that the plots correspond to different parameters, specifically $h_2=2$ for Test Case 2 and $h_2=10$ for Test Case 3, while $h_1 =1$  in both cases. Test Case 3 is considered more challenging due to a higher $\frac{h_2}{h_1}$ ratio, indicating significantly different channel gains or observations. When $\frac{h_2}{h_1}$ is close to one, the two users' observations are more similar, making it easier for the network to generate a reasonable solution. However, in Test Case 3, with a higher $\frac{h_2}{h_1}$ ratio, the task becomes more challenging. In fact, the loss weight parameter $w$ becomes crucial in such scenarios where the channel gains vary significantly.

\subsection{Comparison with Existing Works}\label{sec_comp}

There are a few related works on end-to-end NOMA, including \cite{alberge2018constellation, van2022deep, ninkovic2023weighted, van2020deep}. The most competitive work among these is \cite{ninkovic2023weighted}, which integrates deep learning-based SIC from \cite{van2022deep} with end-to-end NOMA \cite{alberge2018constellation}. However, its network structure assumes both receivers have access to both channel gains, which is impractical. Given its relevance, we specifically compare our results with this work, emphasizing that in our AE-NOMA design, each user knows only its own channel gain.

\subsubsection{BER}
Both \cite{alberge2018constellation} and \cite{ninkovic2023weighted} present BER plots versus $\rm SNR_1$ within the range of $\rm [14, 24]  \, \rm dB$, whereas our plots cover a  more practical SNR range of $\rm [0, 20] \, \rm dB$. Specifically, at ${\rm SNR}_1 = 14 \, \rm dB$, the BER of the weaker user in both \cite{alberge2018constellation} and \cite{ninkovic2023weighted} is approximately  $10^{-1}$, while our results in Fig.~\ref{fig_BER_SNR} and Fig.~\ref{fig_BER_SNRV} demonstrate a BER a little higher than $10^{-3}$ for the same SNR. Notably, these papers lack BER reporting for $\rm SNR_1 < 14,\, \rm dB$, indicating their methods are effective only at high SNRs. Detailed comparison  is provided in Table~\ref{tab:BER}. 

\subsubsection{Loss Function} There are several key reasons for our approach's improvement. First and foremost, our loss function differs fundamentally. Previous works \cite{alberge2018constellation, ninkovic2023weighted, van2022deep} implement SIC  for the stronger user, aligning with information theory principles. In particular, \cite{van2022deep} proposed SICNet, a deep neural network-based implementation of SIC, with \cite{ninkovic2023weighted} following a same structure. Similarly, \cite{alberge2018constellation} incorporates SIC, though  implicitly. If $\mathcal{L}(\mathbf{x},\hat{\mathbf{x}})$ represents the cross-entropy loss between the input $\mathbf{x}$ and output $\hat{\mathbf{x}}$, then the loss function in \cite{alberge2018constellation} is expressed as \(
\mathcal{L}_1 (\mathbf{s},\hat{\mathbf{s}}) + \mathcal{L}_2 (\mathbf{s}_2, \hat{\mathbf{s}}_2)
\)
where $\mathbf{s} = [\mathbf{s}_1 \; \mathbf{s}_2]$. This setup means that the loss for user~1 is based on decoding both users' messages, similar to SIC. However, this approach imposes an unnecessary constraint since user~1 only needs to decode its own message. Additionally, we introduce an adaptive weighted loss function to emphasize reducing errors for the user with higher error rates. We also incorporate residual connections to enhance learning.

\begin{table}[h]
	\caption{The `worse' BER of the proposed AE-NOMA compared to earlier works for different values of $\rm SNR_1$. }\label{tab:BER}
	\centering
	\begin{tabular}{l|ccccc}
		\hline  
		\multicolumn{1}{l}{}  \Tstrut & \multicolumn{4}{c}{\textit{$\rm SNR_1\; (\, \rm dB)$}}   \\ 
		\hline
		\multicolumn{1}{l|}{Method}   \Tstrut  &7.5   & 14    & 16      & 18          \\ \hline
		 \Tstrut Proposed   &  $4\times10^{-2}$  &   $2\times 10^{-3}$    & $2\times 10^{-4}$ &  $7\times 10^{-4}$ \\
		\cite{ninkovic2023weighted} & NA  &  $7\times 10^{-2}$   & $2\times 10^{-2}$ &  $8\times 10^{-3}$ \\
		 \cite{alberge2018constellation}  &  NA
		    &  $8\times 10^{-2}$   & $2\times 10^{-2}$  &   $9\times 10^{-3}$ \\ \hline 
	\end{tabular}\\\vspace{.1cm}
	{\footnotesize{ $\qquad$$\qquad$$\quad$  ``NA" indicates ``not available."  \hfill}}        
\end{table}

\subsubsection{Fairness}
Another  advantage of the proposed architecture in this paper is its high fairness. In NOMA scenarios, it is common to observe significantly different BERs for the two NOMA users. Achieving fairness often involves imposing new constraints on the capacity region of NOMA, as discussed in \cite{timotheou2015fairness}. However, such approaches, while effective in equalizing rates, limit the capacity region and achieve fairness at the cost of degrading the rate for the stronger user, consequently lowering the overall sum-rate. In contrast, our approach embeds fairness into the training and backpropagation process, optimizing the network to improve the worse BER at each iteration. This results in similar BERs for both UE1 and UE2, as evident in Fig.~\ref{fig_BER_SNR} and Fig.~\ref{fig_BER_SNRV_h2_10}. In contrast, both traditional \cite{cejudo2017power, kara2018ber, han2021study, assaf2020exact} and autoencoder-based constellations exhibit a significant gap between the BERs of the two users. BER fairness is crucial, especially when scaling  the number of users, ensuring reliable communication for each user.

\subsection{Insights Obtained}

Theoretically, it is well-established that when the channel gains of two users are close, the performance gain of NOMA diminishes, and with \( h_1 = h_2 \), the capacity region of NOMA collapses to that of OMA.   In practical scenarios where finite-alphabet and finite-length codewords are used—such as QPSK constellations, as opposed to long, capacity-achieving codes—similar channel gains make SIC decoding highly challenging. This can result in BERs even worse than those in OMA \cite{qi2021over}. Additionally, depending on the power allocation coefficient, the two constellations may overlap, resulting in high BERs.
AE-NOMA is able to overcome these challenges and achieve highly competitive results, particularly when \( h_1 \approx h_2 \). This was demonstrated in Fig.~\ref{fig_BER_SNRV}, where AE-NOMA significantly outperforms QPSK-based NOMA. 

\section{Conclusion}\label{sec_con}
We have developed an autoencoder-based downlink NOMA system that creates super-constellations with distinct symbols tailored to varying inter-user interference levels, using a loss function to achieve an SIC-free structure.
The proposed AE-NOMA architecture minimizes BER by jointly designing the transmit/receive autoencoders and optimizing them together. In the proposed architecture, the average power constraint is enforced through a normalization layer, enabling the proposed AE-NOMA to design more efficient symbols that approach the theoretical lower bounds of BER. Simulations confirm that the proposed structure improves BER by over an order of magnitude for the worst-performing user across a wide SNR range, outperforming both existing autoencoder-based schemes and superimposed QPSK constellations.

\textbf{Acknowledgment:} The authors wold like to thank Selma Benouadah for her assistance in regenerating some of the plots.

\end{document}